\documentclass[pra,onecolumn,floatfix,a4paper,superscriptaddress]{revtex4}
\usepackage{bm,color,graphicx,amsmath,txfonts}

\usepackage[colorlinks, citecolor=blue,linkcolor=blue]{hyperref}

\newcommand{\Tr}{{\rm Tr}}

\newcommand{\e}{{e}}

\begin{document}

\title{Quantum Steering vs Entanglement and Extracting Work in an Anisotropic Two-Qubit Heisenberg Model in Presence of External Magnetic Fields with DM and KSEA Interactions}

\author{M. Amazioug} \thanks{amazioug@gmail.com}
\affiliation{LPTHE-Department of Physics, Faculty of sciences, Ibnou Zohr University, Agadir, Morocco.}

\author{M. Daoud}  \thanks{m\_daoud@hotmail.com}
\affiliation{LPMS, Department of Physics, Faculty of Sciences, University Ibn Tofail, K\'enitra, Morocco.}

\begin{abstract}

We examine the Dzyaloshinski-Moriya (DM) and Kaplan-Shekhtman-Entin-Wohlman-Aharony (KSEA) interactions in thermal equilibrium submitted to the anisotropic Heisenberg two-qubit model in an inhomogeneous magnetic field. The steerability between the two qubits is evaluated using quantum steering. The concurrence serves as a witness to quantum entanglement. Both the extracted work and the ideal efficiency of the two qubits are quantified. We discuss how quantum correlations behave in relation to the bath's temperature and the Kaplan-Shekhtman-Entin-Wohlman-Aharony coupling parameter. We find that the nonclassical correlations in a two-qubit Heisenberg XYZ Model are fragile under thermal effects. Nevertheless, the results indicate that the concurrence is stronger than quantum steering under thermal effects. We obtained that an extraction of work is comparable with the bare energies.

\end{abstract}


\maketitle

\section{Introduction}

Entanglement is regarded as a valuable resource in quantum information processing due to its nonclassical nature which enhances the performance of quantum tasks. The study of this resource in quantum communication has received more attention in recent years, for several applications such as quantum teleportation \cite{CHBennett1993}, superdense coding \cite{CHBennett1992}, telecloning \cite{VScarani2005} and quantum cryptography \cite{AKEkert1991}. In 1935, Schr\"odinger proposed the idea of Einstein-Podolsky-Rosen (EPR) steering to discuss the EPR paradox \cite{AEinstein1935,ESchrodinger1935,ESchrodinger1936}. The steerability of the two entangled bipartite states is measured using quantum steering. This quantifier displays the asymmetric property between Alice and Bob, two entangled observers. By taking advantage of their shared entanglement, Alice can thus alter (i.e. "steer") Bob's state \cite{JSchneeloch2013}. It is obvious the minimization of the decoherence effects is necessary in order to take advantage of the quantum features of nonclassical correlations in quantum information processing \cite{WHZurek2003}.\\

Recently, quantum thermodynamics has attracted a special attention \cite{SVinjanampathy2016}. Indeed, the realization of thermodynamic processes at the quantum level is one of the most fascinating issues of this emerging field of research in quantum information \cite{HTQuan2007,HTQuan2006,SWKim2011,TEHumphrey2002,BLeggio2015,MJames2016,AHewgill2018}. 
Among those subjects we quote the maximum entropy production principle \cite{LMMartyushev2006,GPBeretta2009,GPBeretta2014,BMilitello2018} and typicality \cite{SPopescu2006,JGemmer2001}. Numerous findings have recently been made using the theoretical framework of thermodynamic resource theory (TRT). In particular, we mention the generalized second law \cite{FGSLBrandao2015,PCwiklinski2015,MLostaglio2015} and the thermomajorization requirement \cite{MHorodecki2013}. The experimental realization of these protocols may be impossible because these theorems are derived by assuming that extremely complex thermodynamic processes are also admissible. Work extraction protocols play a significant role among all the potential quantum processes \cite{MHorodecki2013,PSkrzypczyk2014}. However, the majority of them are difficult to implement experimentally. For this reason, numerous initiatives have been made to comprehend how to create realizable thermodynamic protocols \cite{MLostaglio2018,CPerry2018}. But for experimental realization, the majority of proposals call for a precise system control. For instance, in a process with various steps, it might be necessary to toggle a particular two-qubits interaction on and off for a period of time unique to each one.\\

The first type of two-qubit interaction we shall consider in this sense is Dzyaloshinsky-Morya (DM). The origin lies on the work of Morya by expanding the Anderson theory of superexchange to incorporate spin-orbit coupling in 1960, resulting in a microscopic theory of anisotropic superexchange interaction. Through the use of perturbation theory, he discovered that the leading anisotropy contribution to the interaction between two nearby spins $\vec\sigma_1$ and $\vec\sigma_2$ writes as
$\mathcal{H}_{\rm DM} = {\bf D}\!\cdot\!({\vec\sigma}_1\times{\vec\sigma}_2)$ \cite{TMOriya19601,TMOriya19602},
where ${\bf D} = (D_x, D_y, D_z)$ is a constant vector that describes a DM interaction and ${\vec\sigma}_i$ denotes the vector of Pauli matrices ${\vec\sigma}_i=(\sigma_i^x,\sigma_i^y,\sigma_i^z)$, where $i=1,2$. The Dzyaloshinsky-Moriya (DM) interaction (spin-orbit coupling) is the name given to the interaction that reproduces Dzyaloshinsky's antisymmetric term. Moriya also discovered the second-order correction term ${\cal H}_{\rm KSEA} = {\vec\sigma}_1\!\cdot\!{\tilde{\rm K}}\!\cdot\!{\vec\sigma}_2$ \cite{TMOriya19601,TMOriya19602}, where $\tilde{\rm K}$ is a symmetric traceless tensor. This interaction was previously thought to be ignored in comparison to the antisymmetric contribution. The O(3) invariance of the isotropic Heisenberg system, which is broken by the DM term, can be restored by the symmetric term, according to Kaplan \cite{TAKaplan1983} and later Shekhtman, Entin-Wohlman, and Aharony \cite{LShekhtman1992,LShekhtman1993}. In literature, various works have taken into account the DM with or without KSEA coupling to investigate the quantum properties of two interacting qubits  \cite{MAYurischev2020,Daoud2023,NHabiballah2018,Huca2022}.\\

In this paper, we investigate the quantum correlations, and especially quantum steering, the anisotropic Heisenberg two-spin-1/2 model in an inhomogeneous magnetic field with both DM and KSEA interactions in thermal equilibrium. We use the concurrence to quantify the amount of entanglement and quantum steering to quantify the steerabilities of the two-qubits. We show that the entanglement and quantum steering depend on the KSEA coupling and the bath temperature. We quantify the extracted work and efficiency. We discuss the influence of the KSEA coupling and the bath temperature on the the extracted work and efficiency.\\

The article is organized as follows. In Section II, we give the model and density matrix of the system under study. In Section III is devoted to the study of quantum correlations between the two-qubits using the concurrence and the quantum steering quantifiers. The extracted work and efficiency is discussed in Section IV. Concluding remarks close this paper.

\section{Model}

The system under consideration consists of two-qubits (qubit 1 + qubit 2) anisotropic Heisenberg XYZ chain in thermal equilibrium at temperature $T$ in the presence of an external magnetic field $b$ along the $z$ axis. We consider that the two qubit are coupled via both types of interactions: Dzyaloshinski-Moriya (DM) interaction which can be writes as
\begin{equation}
   \label{eq:H_DM_Dxyz}
  \mathcal{H}_{\rm DM} = D_x(\sigma_1^y\sigma_2^z-\sigma_1^z\sigma_2^y)
	 +D_y(\sigma_1^z\sigma_2^x-\sigma_1^x\sigma_2^z)
	 +D_z(\sigma_1^x\sigma_2^y-\sigma_1^y\sigma_2^x),
\end{equation}
The KSEA interaction writes as
\begin{eqnarray}
   \label{eq:H_KSEA_Gxyz}
	{\cal H}_{\rm KSEA} = &&(\sigma_1^x,\sigma_1^y,\sigma_1^z)
	 \left(
      \begin{array}{lll}
      0&{\rm K}_z&{\rm K}_y\\
      {\rm K}_z&0&{\rm K}_x\\
      {\rm K}_y&{\rm K}_x&0
      \end{array}
   \right)\!
	 \left(
      \begin{array}{c}
      \sigma_2^x\\
      \sigma_2^y\\
      \sigma_2^z
      \end{array}
   \right)
	 \nonumber\\
	 &&={\rm K}_x(\sigma_1^y\sigma_2^z+\sigma_1^z\sigma_2^y)
	 +{\rm K}_y(\sigma_1^z\sigma_2^x+\sigma_1^x\sigma_2^z)
	 +{\rm K}_z(\sigma_1^x\sigma_2^y+\sigma_1^y\sigma_2^x),
\end{eqnarray}
where ${\rm K}_x$, ${\rm K}_y$, and ${\rm K}_z$ are the KSEA coupling parameters.

The Hamiltonian governing the evolution of the corresponding system (qubit 1 + qubit 2) along the $z$-direction, is given by
\begin{equation} \label{Hzz}
\mathcal{H} = \mathcal{H}_{1} + \mathcal{H}_{2} + \mathcal{H}_{12},
\end{equation}
where
\begin{equation} \label{H}
\mathcal{H}_{1} = b_1\sigma_1^z,\quad \mathcal{H}_{2} = b_2\sigma_2^z,\quad \mathcal{H}_{12} = J_x\sigma_1^x\sigma_2^x + J_y\sigma_1^y\sigma_2^y + J_z\sigma_1^z\sigma_2^z + D_z(\sigma_1^x\sigma_2^y + \sigma_1^y\sigma_2^x) + K_z(\sigma_1^x\sigma_2^y - \sigma_1^y\sigma_2^x),
\end{equation}
where ${\cal H}_{i}$ $(i=1,2)$ is the Zeeman energy, with $b_1$ and $b_2$ are, respectively, the $z$-component of the external fields applied to qubits 1 and 2. The term ${\cal H}_{\rm H} = J_x\sigma_1^x\sigma_2^x + J_y\sigma_1^y\sigma_2^y + J_z\sigma_1^z\sigma_2^z $ is the anisotropic exchange Heisenberg interactions, $(J_x,~J_y,~J_z)$ are the components of Heisenberg exchange interaction, $D_z$ the $z$-component of Dzyaloshinsky vector and $K_z$ the $z$-component in the KSEA interaction. To simplify, we will limit our investigation to situations where the DM and KSEA interactions exist only along the $z$-axis (i.e., $D_x = D_y = 0$ and $K_x = K_y = 0$), respectively. The ferromagnetic phase is occurs for $J_i > 0$ ($i=x,y,z$). In the two-qubit computational basis $\mathcal{B} = \{|00\rangle, |01\rangle, |10\rangle, |11\rangle \}$, the Hamiltonian (\ref{Hzz}) is given by
\begin{equation}
   \label{eq:HXzz}
   {\cal H}=
	 \left(
      \begin{array}{cccc}
      b_1+b_2+J_z& 0 & 0 &J_x-J_y-2i{\rm K}_z\\
      0 & b_1-b_2 -J_z &J_x+J_y+2iD_z& 0\\
      0 &J_x+J_y-2iD_z\ & -b_1+b_2 -J_z & 0\\
      J_x-J_y+2i{\rm K}_z& 0 & 0 & -b_1-b_2 + J_z
      \end{array}
   \right),
\end{equation}
The corresponding eigenvalues are computed as 
\begin{eqnarray}
\lambda_1 &=& J_z - \sqrt{(b_1+b_2)^2+J^2_-+4K_z^2}\quad,\quad \lambda_2 = J_z + \sqrt{(b_1+b_2)^2+J^2_-+4K_z^2},\nonumber \\ 
&&\lambda_3 = -J_z + \sqrt{(b_1-b_2)^2+J^2_++4D_z^2}\quad,\quad \lambda_4 = -J_z + \sqrt{(b_1-b_2)^2+J^2_++4D_z^2},\label{eq:1a} 
\end{eqnarray}
The corresponding eigenstates are given by
\begin{eqnarray}
|\psi_1\rangle &=& -\e^{-i\varphi_1}\sin{(\theta_1/2)}|00\rangle+\cos{(\theta_1/2)}|11\rangle\quad,\quad |\psi_2\rangle = \e^{-i\varphi_1}\cos{(\theta_1/2)}|00\rangle+\sin{(\theta_1/2)}|11\rangle,\nonumber \\ 
&&|\psi_3\rangle = \e^{-i\varphi_2}\cos{(\theta_2/2)}|01\rangle+\sin{(\theta_2/2)}|10\rangle\quad,\quad |\psi_4\rangle = -\e^{-i\varphi_2}\sin{(\theta_2/2)}|01\rangle+\cos{(\theta_2/2)}|10\rangle,\label{eq:1a} 
\end{eqnarray}
where $J_\pm = J_x\pm J_y$, $\tan{\theta_1}=\frac{\sqrt{J_-^2+5K_z^2}}{b_1+b_2}$, $\tan\varphi_1 = \frac{2K_z}{J_-}$, $\tan{\theta_2}=\frac{\sqrt{J_+^2+5D_z^2}}{b_1-b_2}$ and $\tan\varphi_2 = - \frac{2D_z}{J_+}$. The density operator of the system, is given by
\begin{equation} \label{rhozz}
\hat\rho(T)= \frac{\e^{-\beta \mathcal{H}}}{Z_S} = \frac{1}{Z_S}\sum^4_{j=1} \e^{-\beta \lambda_j} |\psi_j\rangle \langle \psi_j|,
\end{equation}
where $Z_S = {\rm Tr }[\e^{-\beta \mathcal{H}}]$ is the partition function of the system
and $\beta=1/(K_B T)$ ($K_B$ is the Boltzmann constant). From the equation (\ref{rhozz}), the thermal state $\hat\rho(T)$ can be cast in the matrix form
\begin{equation}
   \label{rhozz2}
   {\cal \hat\rho}(T)=
	 \left(
      \begin{array}{cccc}
      \hat\rho_{11}& 0 & 0 & \hat\rho_{14}\\
      0 & \hat\rho_{22} & \hat\rho_{23}& 0\\
      0 &\bar\hat\rho_{23} & \hat\rho_{33} & 0\\
      \bar\hat\rho_{14} & 0 & 0 & \hat\rho_{44}
      \end{array}
   \right)
\end{equation}
where 
\begin{equation}
\hat\rho_{11} = \frac{\e^{-\beta J_z}}{Z_S}\left( \e^{\beta\Delta}\sin^2{(\theta_1/2)}+\e^{-\beta\Delta}\cos^2{(\theta_1/2)} \right)\quad,\quad \hat\rho_{22} = \frac{\e^{\beta J_z}}{Z_S}\left( \e^{\beta\Delta'}\cos^2{(\theta_2/2)}+\e^{-\beta\Delta'}\sin^2{(\theta_2/2)} \right),
\end{equation}

\begin{equation}
\hat\rho_{33} = \frac{\e^{\beta J_z}}{Z_S}\left( \e^{\beta\Delta'}\sin^2{(\theta_2/2)}+\e^{-\beta\Delta'}\cos^2{(\theta_2/2)} \right)\quad,\quad \hat\rho_{44} = \frac{\e^{-\beta J_z}}{Z_S}\left( \e^{\beta\Delta}\cos^2{(\theta_1/2)}+\e^{-\beta\Delta}\sin^2{(\theta_1/2)} \right),
\end{equation}
\begin{equation}
\hat\rho_{41} = \bar\hat\rho_{14} = -\frac{\e^{-\beta J_z}}{Z_S}\e^{i\varphi_1}\sin\theta_1\sinh{(\beta\Delta)}\quad,\quad \hat\rho_{32} = \bar\hat\rho_{23} = \frac{\e^{\beta J_z}}{Z_S}\e^{i\varphi_2}\sin\theta_2\sinh{(\beta\Delta')},
\end{equation}

with $Z_S = 2\e^{\beta J_z}\cosh{(\beta \Delta')}+ 2\e^{-\beta J_z}\sinh{(\beta \Delta)}$, $\Delta=\sqrt{(b_1+b_2)^2+J_-^2+4K_z^2}$ and $\Delta'=\sqrt{(b_1-b_2)^2+J_+^2+4D_z^2}$.

\section{Quantum correlation measures}

\subsection{Quantum steering}\label{S1}

In this section, we will quantify the amount of steerability from qubit 1 (Alice: A) to qubit 2 (Bob: B) and from qubit 2 to qubit 1 using EPR-steering. The degree of steerability is quantified based on Alice measurements. It writes as follows \cite{DCavalcanti2016}
\begin{equation}
	\mathcal{S}^{A\to B}=\max\bigg(0,\frac{\mathcal{I}_{AB}-2}{4}\bigg),
\end{equation}
where the factor 4 is introduced to ensure that the steering quantifier is normalized for a system that was originally prepared in Bell states. By exchanging the roles of $A$ and $B$, the possibility of the steering by performing measurements on the subsystem $B$ written as,
\begin{equation}
\mathcal{S}^{B\to A}=\max\bigg(0,\frac{\mathcal{I}_{BA}-2}{4}\bigg),
\end{equation}
The steering asymmetry is given by
\begin{equation}
\Delta_{12} ={\bigg |}\mathcal{S}^{A\to B} - \mathcal{S}^{B\to A}{\bigg |},
\end{equation}
By applying the Pauli spin  operator ($\sigma_x,\sigma_y,\sigma_z$) as measurements, the EPR inequality of steering from $A$ to $B$ writes as
\begin{equation} \label{e1}
\mathcal{I}_{AB}=H(\sigma_x^{B}|\sigma_x^{A})+H(\sigma_y^{B}|\sigma_y^{A})+H(\sigma_z^{B}|\sigma_z^{A})\geq 2
\end{equation}
where $H$ symbol stands for the Shannon entropy and $H(B|A) = H(\rho_{AB}) - H(\rho_A)$ is the conditional Shannon entropy. In the computational basis $\{|00\rangle,\ |01\rangle,\ |10\rangle, |11\rangle\}$, the density matrix for the system under study is given by (\ref{rhozz2}). According to the Pauli matrices $\sigma_x$, $\sigma_y$, and $\sigma_z$ measurements, the expression of EPR steering inequality in Eq (\ref{e1}) is given by \cite{MYAbd-Rabbou2022}
\begin{equation}\label{e3}
	\begin{split}
	\mathcal{I}_{AB}=&(1/2)\sum_{i=1}^{4}\bigg\{P^{AB}_{x_i}\log_2[P^{AB}_{x_i}]+P^{AB}_{y_i}\log_2[P^{AB}_{y_i}]+P^{AB}_{z_i}\log_2[P^{AB}_{z_i}]\bigg\}
\\&-\sum_{i=1}^{2}\bigg\{P^{A}_{x_i}\log_2[P^{A}_{x_i}]+P^{A}_{y_i}\log_2[P^{A}_{y_i}]+P^{A}_{z_i}\log_2[P^{A}_{z_i}]\bigg\}
	\end{split}
\end{equation}
where $P^{AB}_{x_i},\ P^{AB}_{y_i}$, and $P^{AB}_{z_i}$ correspond to the probability distribution of the two-qubits state $\hat\rho$ (see equation (\ref{rhozz2})), are given by

\begin{equation}
	\begin{split}
	&P^{AB}_{x_1}=P^{AB}_{x_2}=(1+2\ Re[\hat\rho_{14}+\hat\rho_{23}])\quad,\quad
	P^{AB}_{x_3}=P^{AB}_{x_4}=(1-2\ Re[\hat\rho_{14}+\hat\rho_{23}]),\\&
	P^{AB}_{y_{1},y_{2}}=1+2\ Re[\hat\rho_{23}-\hat\rho_{14}]\quad,\quad
	P^{AB}_{y_{3},y_{4}}=1-2\ Re[\hat\rho_{23}-\hat\rho_{14}]\quad,\quad
	P^{AB}_{z_i}=4 \hat\rho_{ii}.
	\end{split}
\end{equation}
where $P^{A}_{x_i},\ P^{A}_{y_i}$, and $P^{A}_{z_i}$ are the probability distribution the reduced single qubit state $\hat\rho_{A} $, with

\begin{equation}
	P^{A}_{x_1} = P^{A}_{x_2} = 1,\quad P^{A}_{y_1} = P^{A}_{y_2} = 1, \quad P^{A}_{z_1}=1+(\hat\rho_{11}+\hat\rho_{22}-\hat\rho_{33}-\hat\rho_{44}) \quad \text{and} \quad P^{A}_{z_2}=1-(\hat\rho_{11}+\hat\rho_{22}-\hat\rho_{33}-\hat\rho_{44}).	
\end{equation}
And $\mathcal{I}_{BA}$ quantifies the steering from Bob to Alice. It is given by
\begin{eqnarray}\label{e9}
\mathcal{I}_{BA}&=&(1/2)\sum_{i=1}^{4}\bigg\{P^{AB}_{x_i}\log_2[P^{AB}_{x_i}]+P^{AB}_{y_i}\log_2[P^{AB}_{y_i}]+P^{AB}_{z_i}\log_2[P^{AB}_{z_i}]\bigg\}
\nonumber\\
&-&\sum_{i=1}^{2}\bigg\{P^{B}_{x_i}\log_2[P^{B}_{x_i}]+P^{B}_{y_i}\log_2[P^{B}_{y_i}]+P^{B}_{z_i}\log_2[P^{B}_{z_i}]\bigg\},
\end{eqnarray}
where,
\begin{eqnarray}
P^{B}_{x_1} = P^{B}_{x_2}&=& 1,\quad P^{B}_{y_1} = P^{B}_{y_2}= 1 , \quad P^{B}_{z_1}=1+(\hat\rho_{11}-\hat\rho_{22}+\hat\rho_{33}-\hat\rho_{44}) \quad \text{and} \quad P^{B}_{z_2}=1-(\hat\rho_{11}-\hat\rho_{22}+\hat\rho_{33}-\hat\rho_{44}).
\end{eqnarray}
There are various possibilities of steerability between the qubit 1 on Alice hand $(A)$ and qubit 2 on Bob hand $(B)$:  (*) if $\Delta_{12}>0$ (one-way steering), i.e, $\mathcal{S}^{A\to B}>0$ and $\mathcal{S}^{B\to A}=0$ or $\mathcal{S}^{A\to B}=0$ and $\mathcal{S}^{B\to A}>0$, (**) if $\Delta_{12}=0$ (two or no-way steering), i.e, $\mathcal{S}^{A\to B}=\mathcal{S}^{B\to A}=0$ (qubit 1 can’t steer qubit 2 and vice versa even if they are entangled), or $\mathcal{S}^{A\to B}=\mathcal{S}^{B\to A}>0$ (qubit 1 can steer qubit 2 and vice versa). In fact, a steerable state is always not separable while an entangled state is not always steerable.

\begin{figure}[!htb]
\minipage{0.32\textwidth}
  \includegraphics[width=\linewidth]{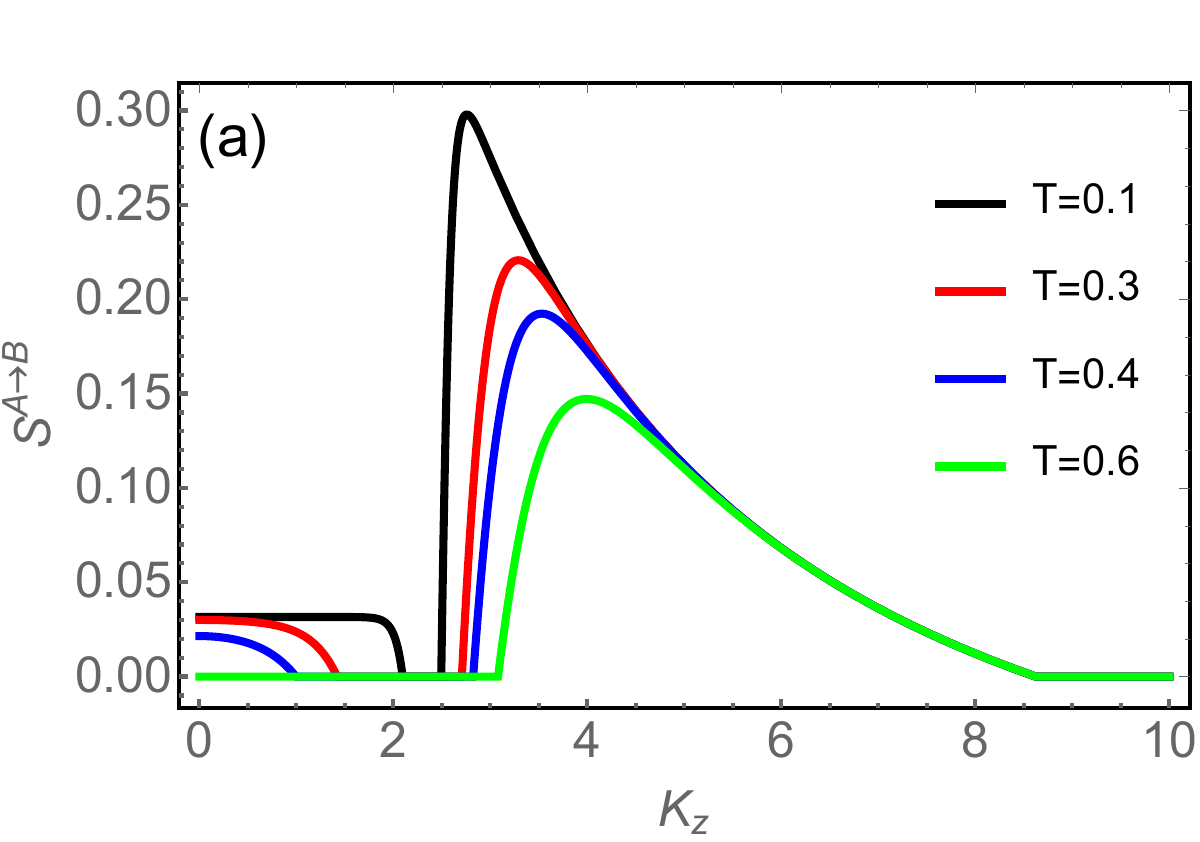}
\endminipage\hfill
\minipage{0.32\textwidth}
  \includegraphics[width=\linewidth]{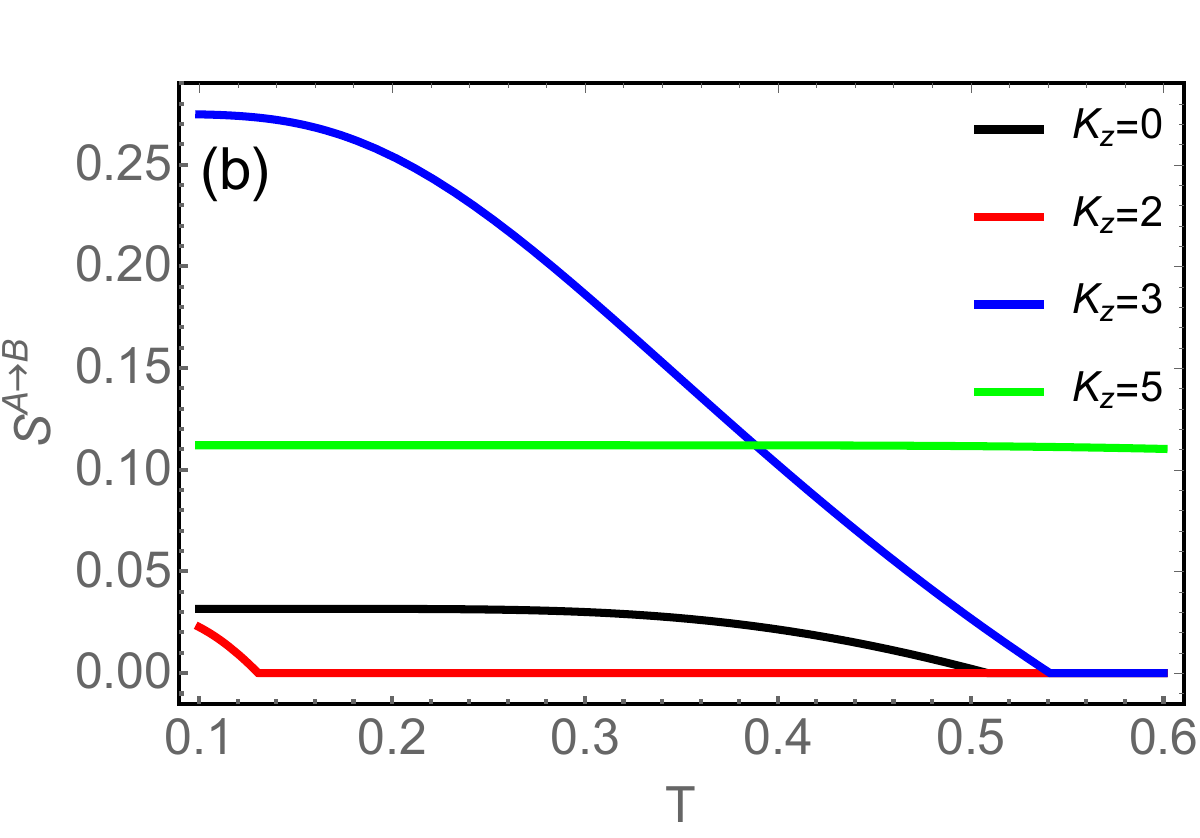}
\endminipage\hfill
\minipage{0.32\textwidth}
  \includegraphics[width=\linewidth]{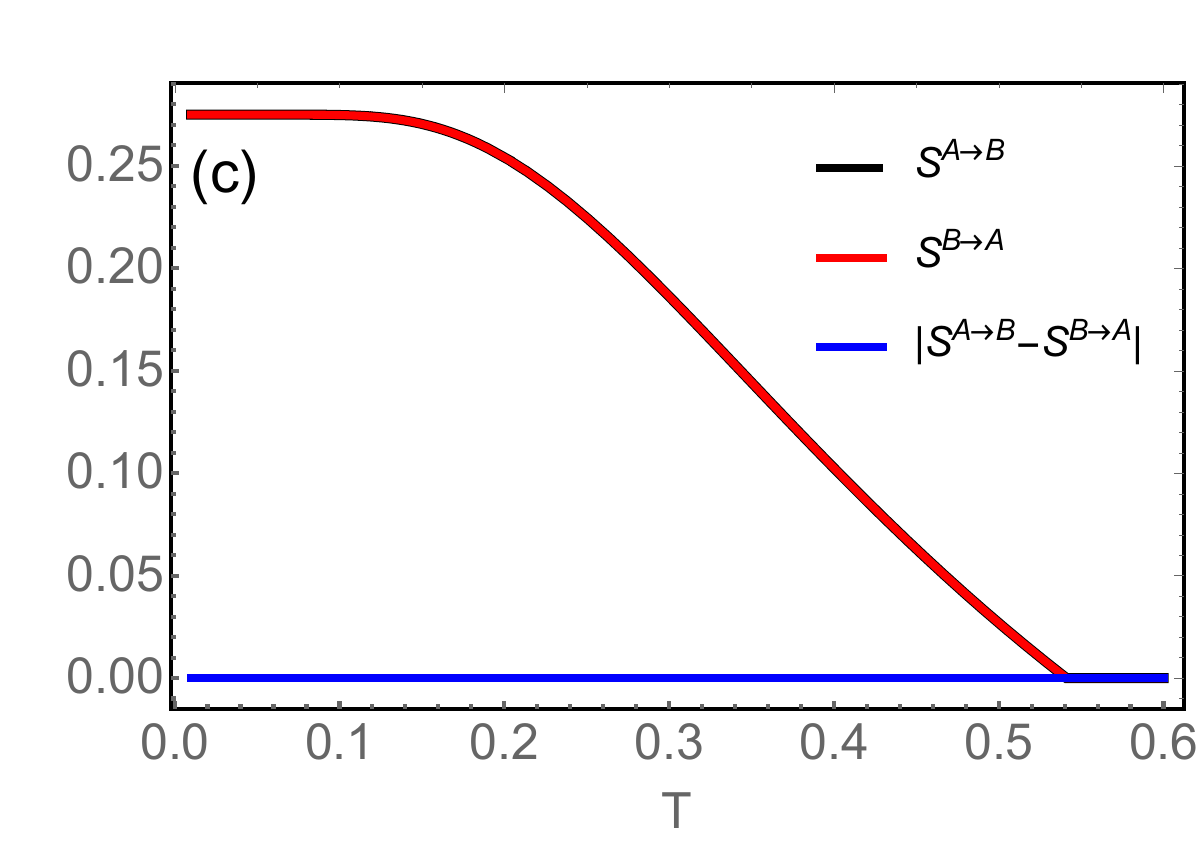}
\endminipage\hfill
\caption{Plot of the quantum steering $\mathcal{S}^{A\to B}$ from qubit $A$ to qubit $B$, (a) versus the KSEA coupling constant $K_z$ for different values of the bath temperature $T$, (b) versus the bath temperature $T$ for various values of $K_z$, (c) comparison of the quantum steering $S^{A\to B}$ and $S^{B\to A}$ for $K_z=3$. The values of additional parameters are $b_1 = 2$, $b_2 = 1$, $J_y=2$, $J_x=J_z=2$ and $D_z=1$.}
\label{Sab}
\end{figure}
In the Fig. \ref{Sab}(a) we plot steerability $\mathcal{S}^{A\to B}$ versus the coupling-constant $K_z$ for various values of the temperature $T$. This figure displays that qubit 1 steer qubit 2 when $K_z\leq 2$ and $T\leq 0.4$. This steerabilities decreases to vanish when $K_z\approx 2$. Moreover, we observe that $\mathcal{S}^{A\to B}$ exhibits a sudden birth and decreases after achieving its maximum value. When $K_z\geq 5$, the quantity $\mathcal{S}^{A\to B}$ is independent of the temperature as it is shown in Fig. \ref{Sab}(a). Besides, the steerabilities $\mathcal{S}^{A\to B}$ is influenced by the temperature, i.e., $\mathcal{S}^{A\to B}$ decreases when we increase the temperature. This can be explained by decoherence effects induced by the increase of the temperature.

In Fig. \ref{Sab}(b) we plot steerability $\mathcal{S}^{A\to B}$ versus $T$ for various values of $K_z$. We remark that $\mathcal{S}^{A\to B}$ decreases for all values of parameter $K_z$, while it remains unchanged under the temperature $T$.

We plot in Fig. \ref{Sab}(c) the steearbilities $\mathcal{S}^{A\to B}$ , $\mathcal{S}^{B\to A}$ and the asymmetry $\Delta_{12}={\bigg |}\mathcal{S}^{A\to B}-\mathcal{S}^{B\to A}{\bigg |}$. We note that $\mathcal{S}^{A\to B}=\mathcal{S}^{B\to A}>0$ (i.e. $\Delta_{12}=0$) when $T\leq 0.13$. This witnesses the existence of two-way steering between the qubit 1 and qubit 2. Moreover, when $T\geq 0.13$ we have $\mathcal{S}^{A\to B}=\mathcal{S}^{B\to A}=0$. This means no-way steering between the qubit 1 and the qubit 2 (i.e. $\Delta_{12}=0$). 

\subsection{Quantum entanglement}\label{EoF}
 
In this subsection, we will introduce the entanglement between the qubit 1 and the qubit 2 and its relation to quantum steering. We use the concurrence to quantify the entanglement between the two-qubit \cite{WKWootters1998}. The concurrence is defined for the bipartite X form state (see equation (\ref{rhozz2})), as
\begin{equation}
\mathcal{C}(\hat\rho) = 2\text{max}\bigg\{0,~~|\hat\rho_{14}|-\sqrt{\hat\rho_{33}\hat\rho_{22}},~~|\hat\rho_{23}|-\sqrt{\hat\rho_{44}\hat\rho_{11}}\bigg\}
\end{equation}
It is important to note that the concurrence $\mathcal{C}(\hat\rho)$ ranges from 0 to 1; the minimal and maximal values represent, respectively, separable and maximally entangled states.
\begin{figure}[!htb]
\minipage{0.32\textwidth}
  \includegraphics[width=\linewidth]{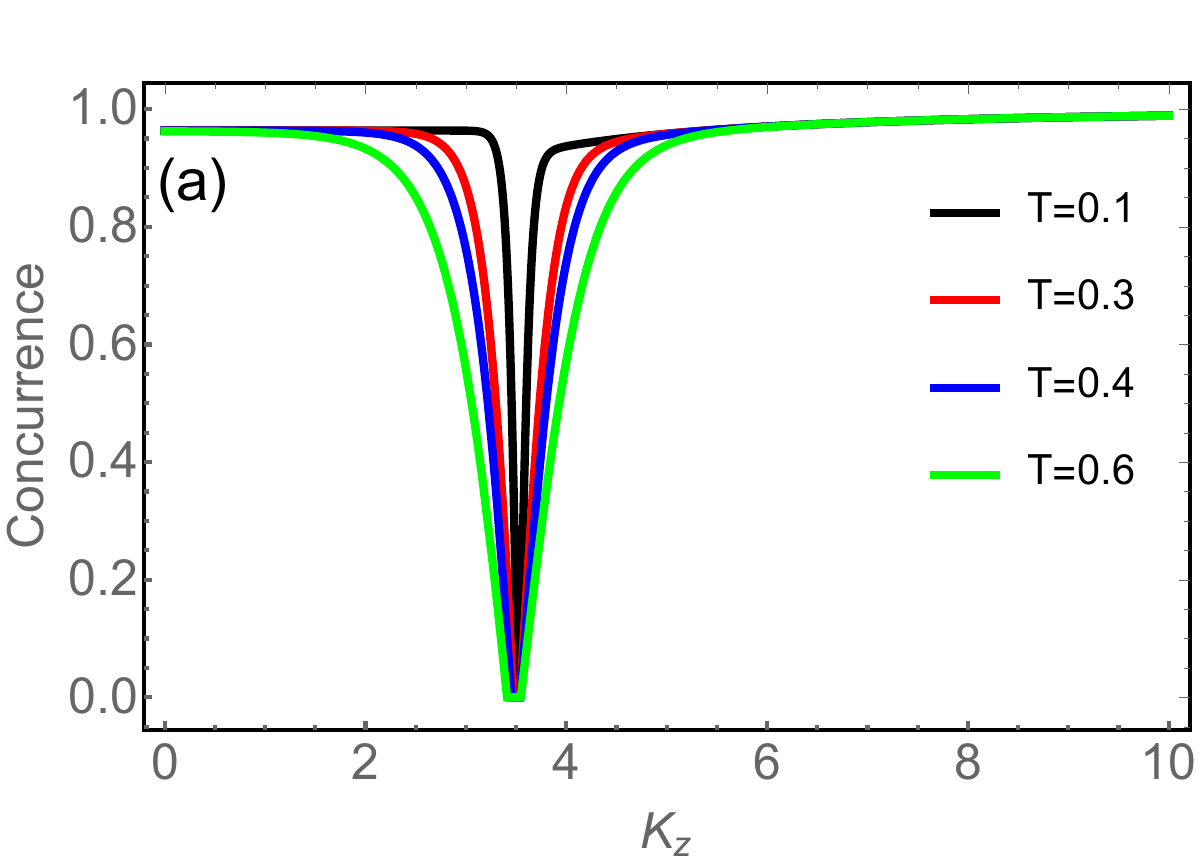}
\endminipage\hfill
\minipage{0.32\textwidth}
  \includegraphics[width=\linewidth]{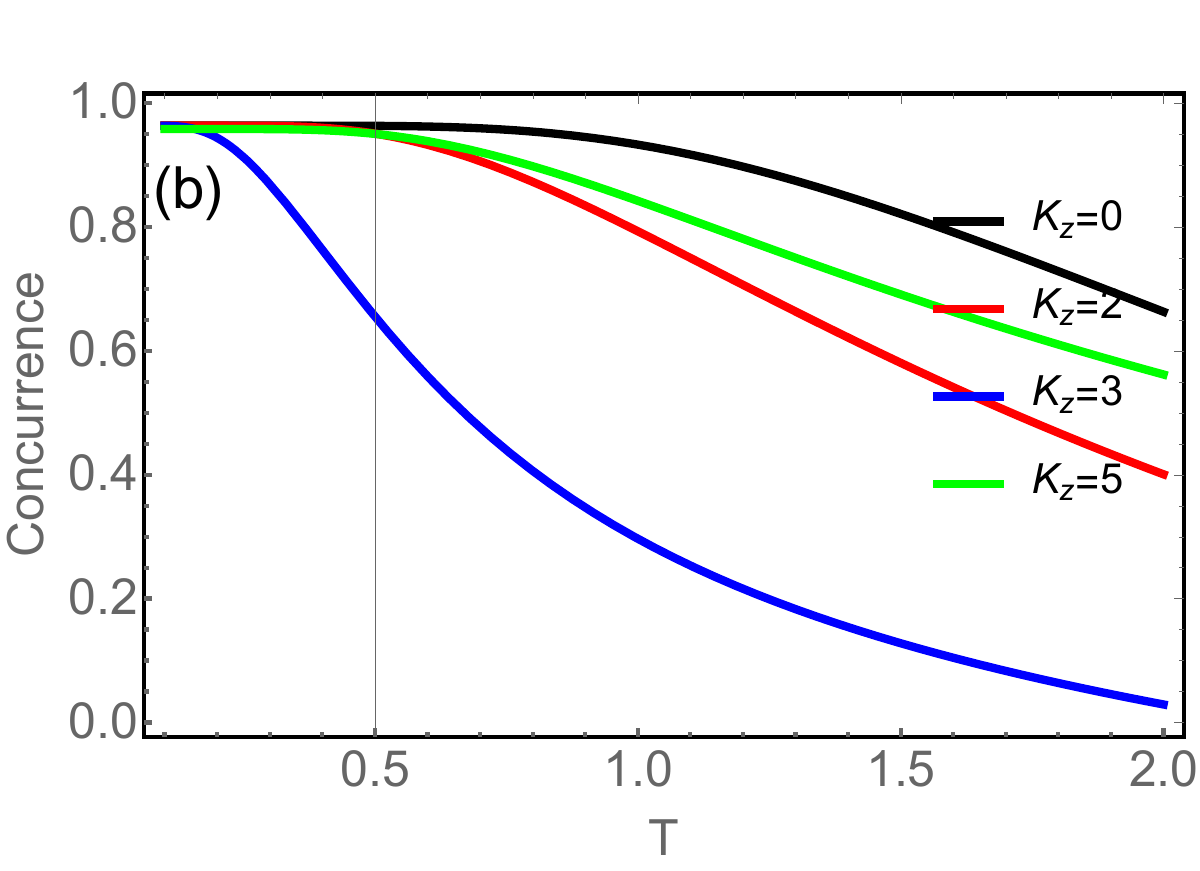}
\endminipage\hfill
\minipage{0.32\textwidth}
  \includegraphics[width=\linewidth]{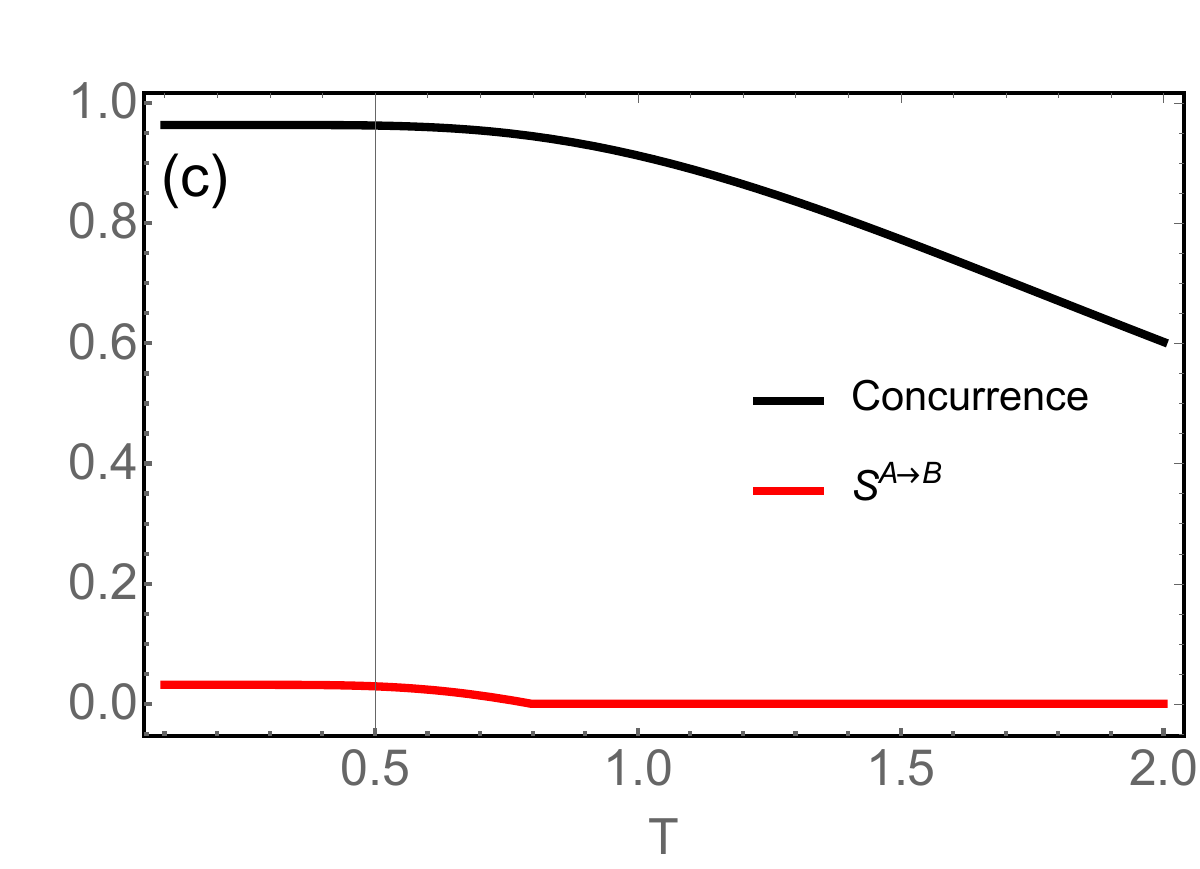}
\endminipage\hfill
\caption{Plot of the concurrence between the two-qubit, (a) versus the KSEA coupling constant $K_z$ for different values of the bath temperature $T$, (b) versus the bath temperature $T$ for various values of $K_z$, (c) comparison of the concurrence and quantum steering $S^{A\to B}$ for $K_z=1$. The values of additional parameters are $b_1 = 2$, $b_2 = 1$, $J_x=1$, $J_y=J_z=2$ and $D_z=1$.}
\label{C}
\end{figure}

We plot in Fig. \ref{C}(a) the concurrence $\mathcal{C}$ versus the $z$-component of the KSEA interaction $K_x$ for different values of the temperature $T$. This figure displays differents dips of concurrence for each value of the temperature $T$ when $2\leq K_z\leq 5.5$. Moreover, when $0\leq K_z\leq 2$ and $6\leq K_z\leq 10$, the concurrence tends to reach its maximum value. This maximum value remains constant even when the temperature increases as depicted in Fig. \ref{C}(a). 

In Fig. \ref{C}(b) we display the concurrence $\mathcal{C}$ versus the temperature $T$ for different values of the $z$-component of the KSEA interaction $K_x$. We note that the entanglement between the two qubits decreases under the influence of the thermal effects (decoherence phenomenon).

We plot in Fig. \ref{C}(c) the concurrence $\mathcal{C}$ and steerability $\mathcal{S}^{A\to B}$ versus the temperature $T$. This figure shows that the steering measure remains bounded by the entanglement (concurrence). We observe again that the concurrence $\mathcal{C}$ and steering $\mathcal{S}^{A\to B}$ decreases following the same behavior. According to Fig. \ref{C}(c), we can see that when $T>0.75$, the concurrence $\mathcal{C}$ is positive ($\mathcal{C}>0$), while the steering $\mathcal{S}^{A\to B}$ vanishes. In contrast, when $T \leq 0.75$, the concurrence $\mathcal{C}>0$ and also the steering $\mathcal{S}^{A\to B}>0$. This implies that the steerable state must be entangled even though the entangled state is not always steerable.

\section{quantum thermodynamics}

In this section, we will study the extracted work of the system consisting of two-qubits 1 and 2. We followed the schematics of the thermalization protocol phases in the presence of a thermal bath, investigated in \cite{Bellomo2019}. At time $t_1$, there are no interactions between two-qubits. The interaction is then activated, and they thermalize simultaneously (from $t_2$ to $t_3$). These two-qubits are thermalize together from time $t_2$ to time $t_3$ ($t_3 - t_2 \gg \tau_r$, where $\tau_r$ represents the typical evolution time for the system at this phase). The extracted work is quantified through the partition functions of the system $Z_S$, ($\hbar = K_B = 1$) \cite{Bellomo2019}

\begin{equation} \label{H}
W = T \ln\left( \frac{Z_1Z_2}{Z_S}\right) - \langle \mathcal{H}_{12} \rangle_{t_3},
\end{equation}
where $Z_{1(2)} = \Tr[\e^{\beta \mathcal{H}_{1(2)}}]$ is the partition function of system $1(2)$, $\langle \mathcal{H}_{12} \rangle_{t_3} = \Tr[\mathcal{H}\hat\rho (T)] - \Tr[(\mathcal{H}_1+\mathcal{H}_2)\hat\rho (T)]$, $\hat\rho (T)=\e^{-\beta \mathcal{H}}/Z_S$, $Z_S=\Tr[\e^{-\beta \mathcal{H}}]$ and $\beta=1/T$. The ideal efficiency is quantified as \cite{Bellomo2019}
\begin{equation}
\eta = \frac{W}{\bigg(-\Delta\langle \mathcal{H}_{12} \rangle \bigg)}\leq 1
\end{equation}
where $\bigg (-\Delta \langle \mathcal{H}_{12} \rangle \bigg) = \langle \mathcal{H}_{12} \rangle_{t_2}-\langle \mathcal{H}_{12}\rangle_{t_3}$ with $\langle \mathcal{H}_{12} \rangle_{t_2}=0$, is the minimal amount of work required by the system to complete one cycle. The global entropic term writes as
\begin{equation}
S_G = T\bigg(S(\hat\rho (T))-S\left(\hat\rho_1(T)\otimes \hat\rho_2(T)\right)\bigg),
\end{equation}
where $\hat\rho_{1(2)}(T)=\e^{-\beta \mathcal{H}_{1(2)}}/Z_{1(2)}$ and $S(\hat\rho (T)) = - \Tr[\hat\rho (T) \ln(\hat\rho (T))]$. The energy difference is given by 
\begin{equation}
E_d = W + S_G,
\end{equation}
The local entropic term is writes as
\begin{equation}
S_l = T \bigg (S(\hat\rho_1^{r} (T)\otimes \hat\rho_2^{r}(T)) - S(\hat\rho_1 (T)\otimes \hat\rho_2 (T))\bigg ),
\end{equation}
where $\hat\rho_{1(2)}^{r} (T) = \Tr_{2(1)}[\hat\rho (T)]$. The local work is given by
\begin{equation}
W_l = W - TS(1:2),
\end{equation}
where $S(1:2) = S(\hat\rho_1^{r}(T)) + S(\hat\rho_2^{r}(T)) - S(\hat\rho (T))$.

\begin{figure}[!htb]
\minipage{0.32\textwidth}
  \includegraphics[width=\linewidth]{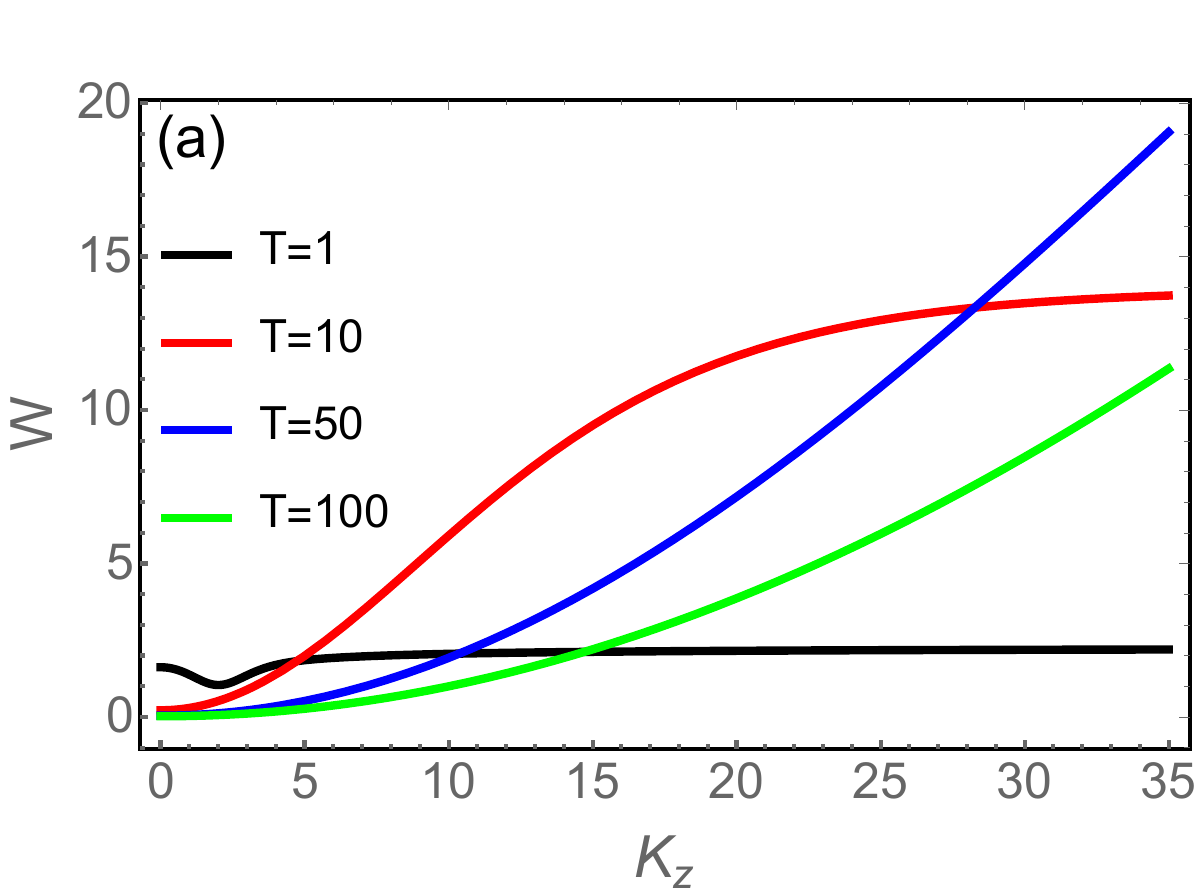}
\endminipage\hfill
\minipage{0.32\textwidth}
  \includegraphics[width=\linewidth]{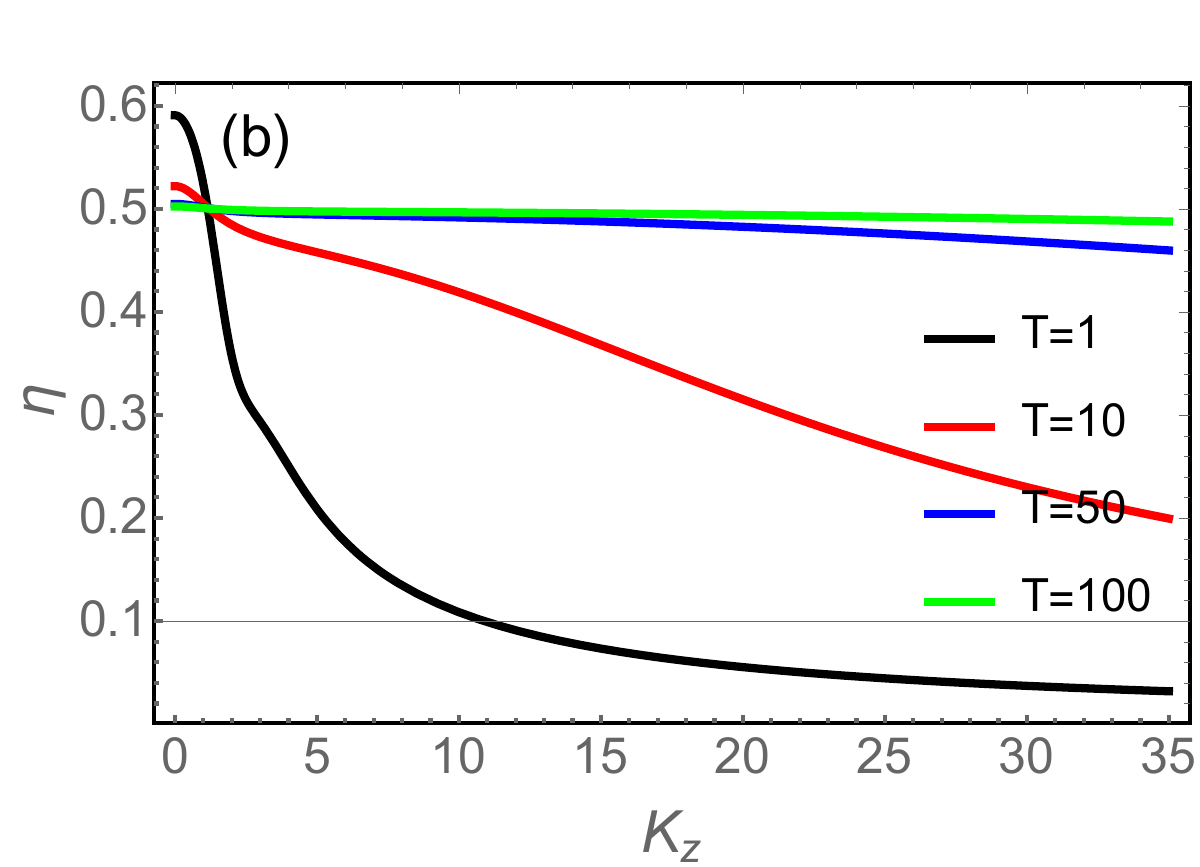}
\endminipage\hfill
\minipage{0.32\textwidth}
  \includegraphics[width=\linewidth]{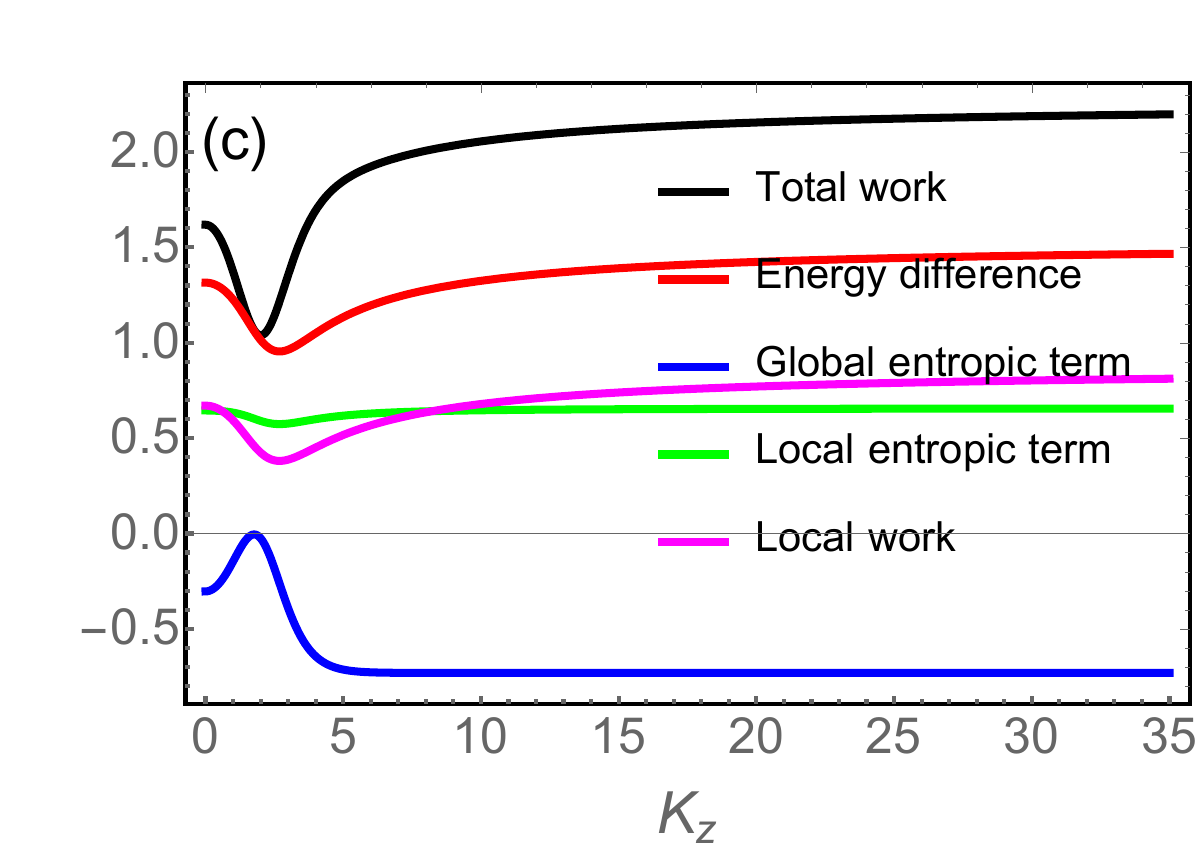}
\endminipage\hfill
\caption{Plot of (a) extracted work $W$ and (b) efficiency $\eta$ of the two-qubit as a function of the KSEA coupling constant $K_z$ for different values of the bath temperature $T$, (c) comparison of different quantities for $T=1$. The values of additional parameters are $b_1 = b_2 = 1$, $J_x=J_z=1$, $J_y=0$ and $D_z=1$.}
\label{W1}
\end{figure}
We plot in Fig. \ref{W1}(a-b) the extracted work $W$ and efficiency $\eta$ versus $K_z$ for various values of the bath temperature $T$. We notice that for every temperature ($T>10$) increasing $K_z$ monotonically increases the extracted work as depicted in Fig. \ref{W1}(a). With respect to the low temperature like $T=1$, the extracted work apparently is not affected by the increase of $K_z$. This shows a direct relationship between the extracted work and KSEA coupling at high temperatures. In contrast, the efficiency remains less than 1 ($\eta < 1$), and decreases with increasing $K_z$ for each temperature value as illustrate in Fig. \ref{W1}(b). The efficiency decreases quickly for $T=1$, compared to other values of $T$. We remark, that at high temperature the efficiency remains approximately 0.5.

In Fig. \ref{W1}(c), we plot the comparison of different quantities (total work, energy difference, global entropic term, local entropic term and local work) as a function of KSEA coupling $K_z$ for a fixed temperature $T=1$. We remark, that all energies are positive, while global entropy term is negative. The maximal extraction of total work $W$ is obtained for $K_z=35$, as depicted in Fig. \ref{W1}(c). For $K_z\geq 3$, we observe that approximately all energies increase to reach their maximum values, while, the global entropy term degrades rapidly. For $K_z\leq 5$, the different energies undergo a small influence in the form of the dips, while, the global entropy term in the form of peak.
\begin{figure}[!htb]
\minipage{0.32\textwidth}
  \includegraphics[width=\linewidth]{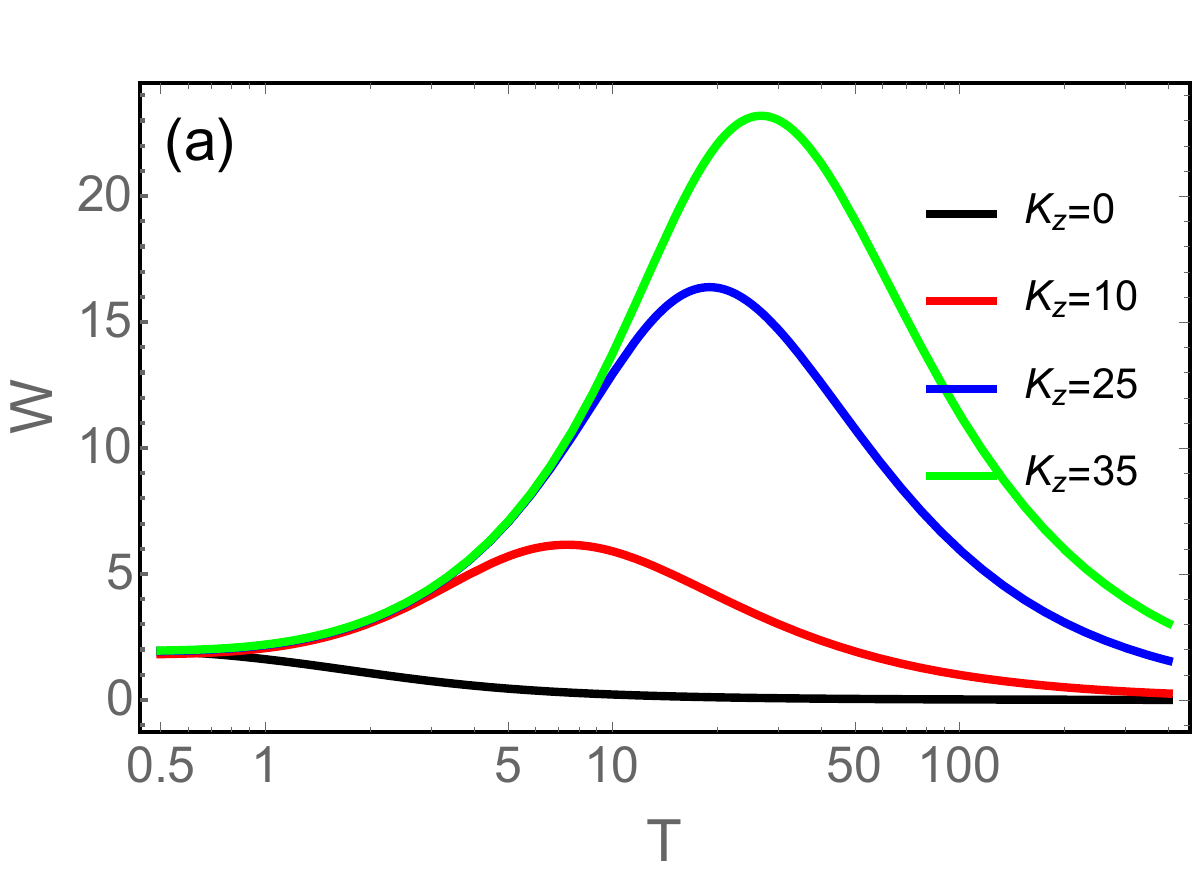}
\endminipage\hfill
\minipage{0.32\textwidth}
  \includegraphics[width=\linewidth]{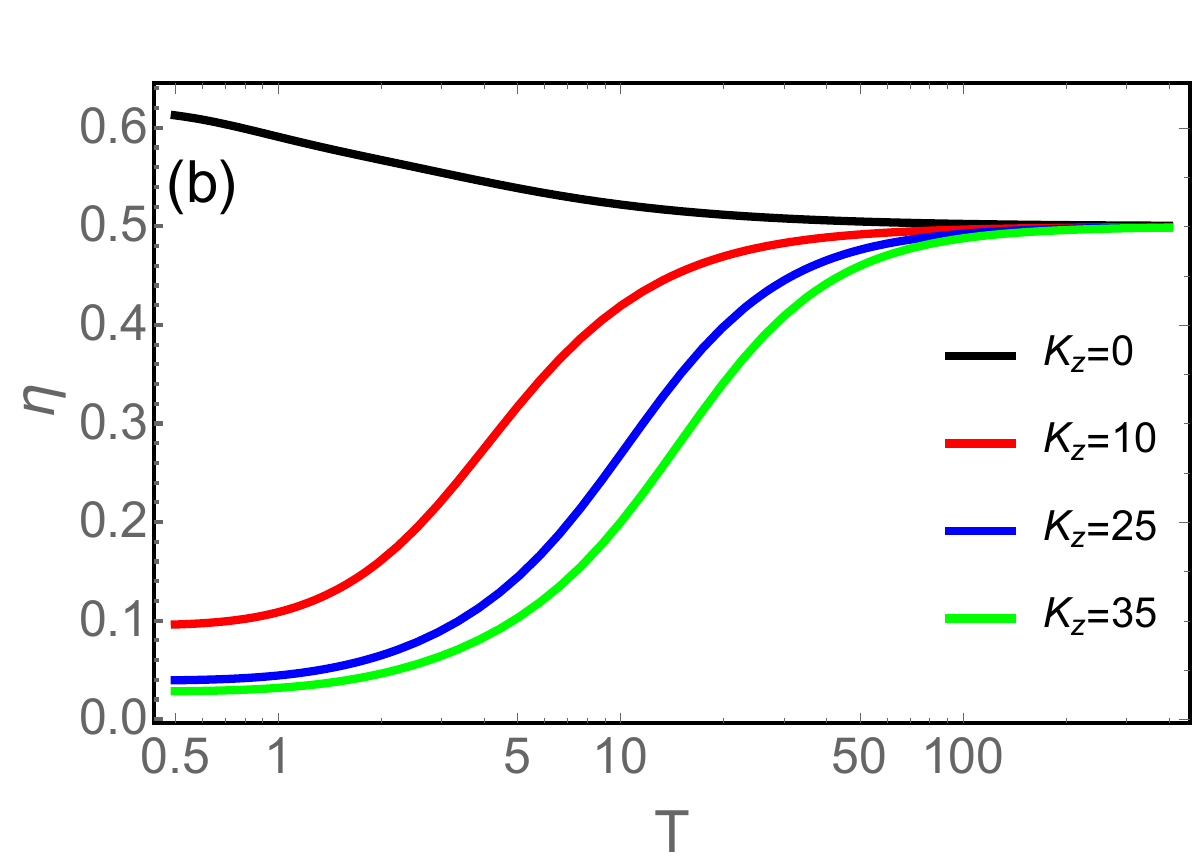}
\endminipage\hfill
\minipage{0.32\textwidth}
  \includegraphics[width=\linewidth]{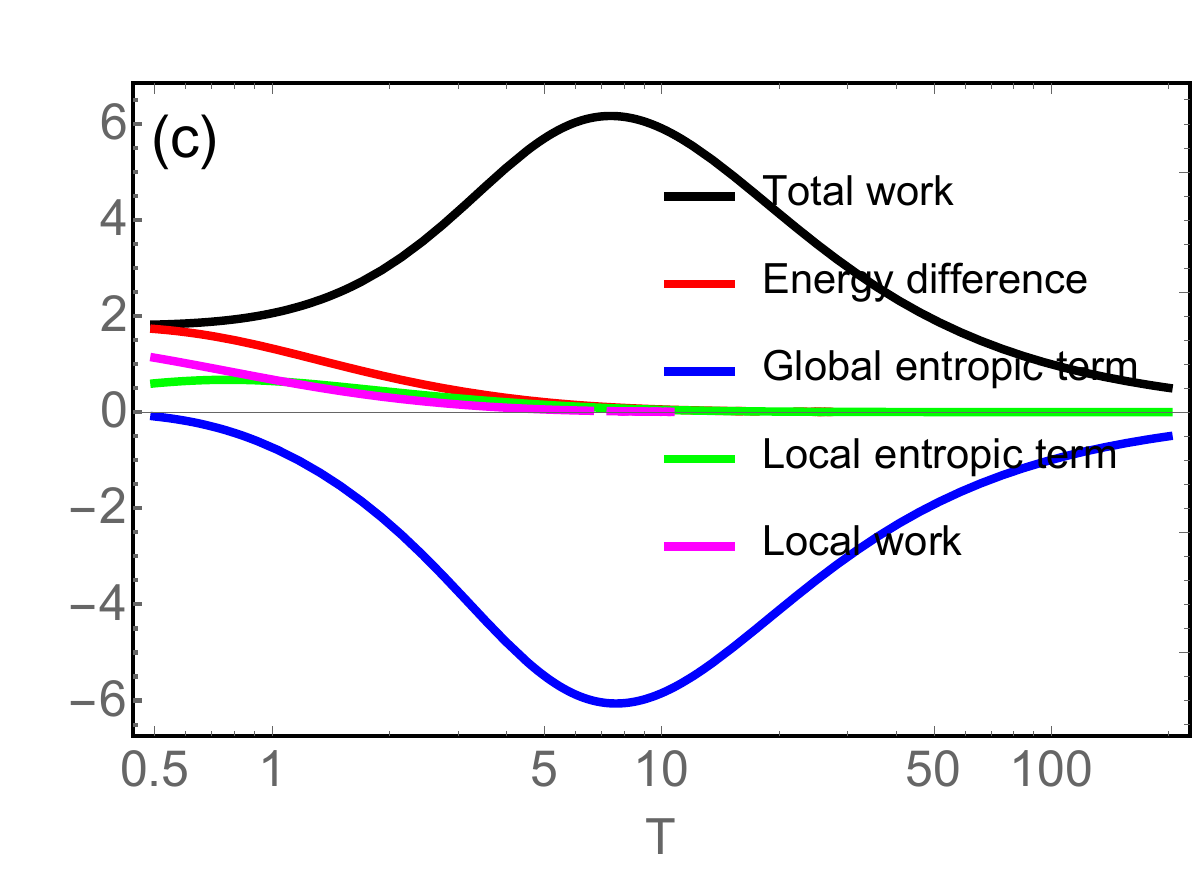}
\endminipage\hfill
\caption{Plot of (a) extracted work $W$ and (b) efficiency $\eta$ of the two-qubit as a function of the bath temperature $T$ for different values of the KSEA coupling constant, (c) comparison of different quantities for $K_z=10$. The values of additional parameters are $b_1 = b_2 = 1$, $J_x=J_z=1$, $J_y=0$ and $D_z=1$.}
\label{W2}
\end{figure}

We plot in Fig. \ref{W2}(a-b) the extracted work $W$ and efficiency $\eta$ versus the bath temperature $T$ for various values of the KSEA coupling $K_z$. This is the most interesting feature of our model is that, given a value of $K_z$, the maximal extraction of work is obtained for a value of temperature such that $T=50$. The extracted work decreases with increasing $T$ for $K_x=0$. Moreover, the work $W$ increases under thermal effects for other values of $K_z$, and decreases after reaching its maximum value at high temperatures, as shown in Fig. \ref{W2}(a). We remark, that the efficiency is always less than 1 ($\eta < 1$). For high temperature ($T\geq 100$), the efficiency tends to be constant at 0.5 ($\eta = 0.5$). Besides, the efficiency increases for all values of KSEA coupling $K_z$, except for $K_z=0$, the efficiency is degraded as implemented in Fig. \ref{W2}(b). The results depicted in figure \ref{W2}, corroborate the results discussed in \cite{Bellomo2019}.

We plot in Fig. \ref{W2}(c), the extracted work and other relevant quantities versus the temperature $T$, for $K_z=10$. For $T=50$, the maximum amount of total work $W$ is achieved. Around that temperature $(T=50)$, we observe that there is a significant difference between the total work $W$ and the local work $W_l$, and the local work quickly decreases to vanishes. The inequality $W_l \leq W$ is a well-known thermodynamic result \cite{KMaruyama2009}, indicating that the presence of correlations in the final thermal state benefits the amount of extracted work. A significant portion of the work is stored in the nonlocal entropic term (mutual information). This implies that the entropic term is responsible for the peak of work extraction. Regarding the total extraction, we can also observe how crucial the global entropic term is in that temperature region ($T=50$). We remark, that the global entropic term is symmetric with total work, as depicted in Fig. \ref{W2}(c). On the local level, the local entropic term, in the high-temperature region, always reduce the total work extracted to zero. The quantitative difference between the local and total extracted work is explained by this difference in behavior between the local and nonlocal parts of the entropy \cite{Bellomo2019}.\\

\section{Conclusion} \label{Conc}

In summary, the quantum correlations, the extracted work  and the efficiency in a two-qubit Heisenberg XYZ model in presence of an inhomogeneous magnetic field with both DM and KSEA interactions in thermal equilibrium are investigated. We have quantified quantum entanglement and quantum steering of the two-qubit. We have discussed and compared two different quantum correlations (concurrence and quantum steering). We have found as expected that for increasing values of temperature, steering and entanglement decrease (the decoherence phenomenon). The enhancement of quantum steering and entanglement in presence of KSEA coupling is discussed. We have proved that quantum entanglement remains more persistent than quantum steering, i.e., the steerable modes are strictly entangled but the entangled modes are not always steerable. We have evaluated the extracted work and efficiency of the two-qubit system under consideration. We have investigated the evolution of extracted work and efficiency as a function of temperature and KSEA coupling. The variation of the extracted work under thermal effects and KSEA coupling, with respect to other energies is discussed.

\section*{Data availability statement}

No Data associated in the manuscript.

\end{document}